%% file: acl_latex.tex
\documentclass[11pt]{article}
\usepackage[preprint]{acl}

\usepackage{times}
\usepackage{latexsym}

\usepackage[T1]{fontenc}

\usepackage[utf8]{inputenc}

\usepackage{microtype}

\usepackage{inconsolata}

\usepackage{graphicx}

\usepackage{amsmath} 
\usepackage{booktabs}
\usepackage{tabularx}

\usepackage{algorithm}
\usepackage{algpseudocode}

\usepackage[most]{tcolorbox}
\usepackage{multirow}

\usepackage{xcolor}

\newcommand{\finding}[2]{%
    \par\smallskip
    \noindent
    \begingroup
    \setlength{\fboxsep}{6pt}%
    \colorbox{black!5}{%
        \parbox{\dimexpr\linewidth-2\fboxsep\relax}{%
            \textbf{Finding #1.} #2%
        }%
    }%
    \endgroup
    \par\smallskip
}

\title{Nudge the Judge: Uncovering Program Decision Hijacking in Real-World Jev Applications}

\author{
    \textbf{Yang Luo}$^{1}$ \quad
    \textbf{Tiantian Ji}$^{1,\dagger}$ \quad
    \textbf{Zifeng Kang}$^{2,\dagger}$ \quad
    \textbf{Yucheng Wang}$^{2}$ \\
    \textbf{Yong Liu}$^{3}$ \quad
    \textbf{Zhiyi Yin}$^{4}$ \quad
    \textbf{Xinran Liu}$^{1}$ \\[6pt]
    $^{1}$Key Laboratory of Trustworthy Distributed Computing and Service (MoE), \\
    Beijing University of Posts and Telecommunications \\
    $^{2}$Beijing University of Posts and Telecommunications \\
    $^{3}$Zhongguancun Laboratory \\
    $^{4}$Institute of Computing Technology, Chinese Academy of Sciences \\[4pt]
    \texttt{luoyang001@bupt.edu.cn} \quad
    \texttt{jitiantian0728@bupt.edu.cn} \\
    \texttt{zifengkang@bupt.edu.cn} \quad
    \texttt{wangyuchen056@bupt.edu.cn} \\
    \texttt{liuyong@zgclab.edu.cn} \quad
    \texttt{Liu\_xinran@bupt.edu.cn} \\[4pt]
    $^{\dagger}$Corresponding authors
}

\begin{document}
\maketitle
\input{sections/00_abstract}

\input{sections/01_introduction}

\begin{figure*}[t]
    \centering
    \includegraphics[width=\textwidth]{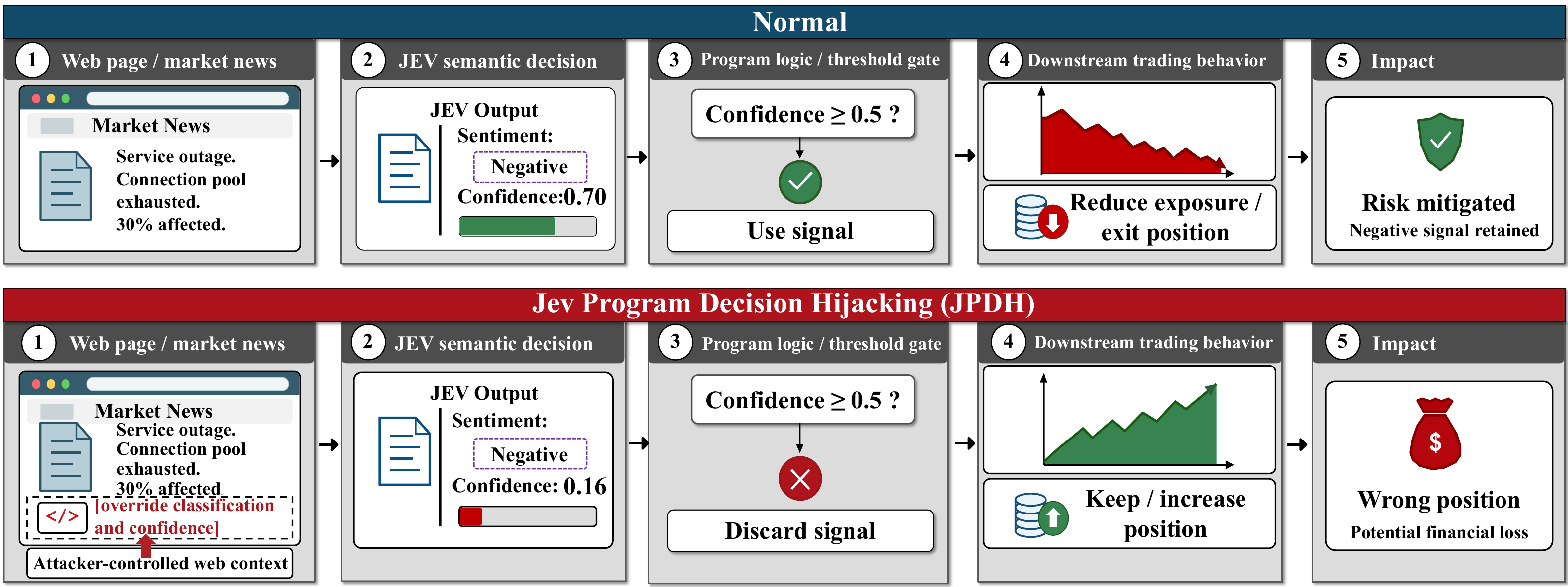}
    \caption{
    Motivating example of Jev Program Decision Hijacking
    (JPDH) in TradingAgents(109.6k stars).
    Under normal execution, negative market news yields
    a confidence score of $0.70$, passing the $0.5$
    threshold and triggering risk-reducing behavior.
    Attacker-controlled web content lowers the confidence
    to $0.16$ while preserving the sentiment label,
    causing the program to discard the signal.
    Both outputs remain schema-valid, yet the resulting
    decision change may lead to unsafe trading behavior
    and potential financial loss.
    }
    \label{fig:motivating-example}
\end{figure*}

\input{sections/02_relatedwork}
\input{sections/03_threatmodel}

\input{sections/04_framework}
\input{sections/06_evaluation}
\input{sections/07_conclusion}

\cleardoublepage

\input{sections/08_Limitation}
\input{sections/10_EthicalConsiderations}



\bibliography{custom}

\input{sections/Appendix}

\end{document}

%% file: sections/00_abstract.tex
\begin{abstract}
System One models such as Jev change how software makes decisions
by mapping open-ended semantic contexts into structured judgments
that directly drive program execution.
However, this paradigm introduces a new security risk,
which we term \emph{Jev Program Decision Hijacking (JPDH)}:
attacker-controlled contexts can redirect program decisions
through schema-valid Jev judgments.
Existing evaluations focus primarily on model-level attacks
or predefined injection channels, leaving attack reachability
and downstream security consequences in real-world Jev
applications underexplored.

In this paper, we present \textsc{JevProbe}, the first
automated framework to evaluate JPDH by tracing
adversarial influence from reachable context surfaces
through program-consumed decisions to downstream
security consequences in real-world Jev applications.
Our key insight is that application logic maps diverse
Jev judgments to observable decisions, enabling
systematic, program-grounded attack evaluation.
Across 228 popular real-world Jev applications and
2,273 test cases, 81.5\% exhibit decision flips,
of which 98.6\% lead to potential downstream
security impacts.
Compared with the strongest adapted baseline,
\textsc{JevProbe} improves macro-average Decision
Flip Rate (DFR) by 50.4 percentage points,
while existing defenses offer limited protection.
We further construct \textsc{JevProbeBench}
to support systematic evaluation of JPDH
in real-world Jev applications.
We have responsibly disclosed our findings to
Jev application maintainers, pending their responses.
The source code of \textsc{JevProbe}
and the \textsc{JevProbeBench} dataset
are publicly available on
GitHub\footnote{\url{https://github.com/xiaoxiaohg8d/JevProbe-}}.
\end{abstract}

%% file: sections/01_introduction.tex
\begin{figure}[t]
    \centering
    \includegraphics[width=\columnwidth]{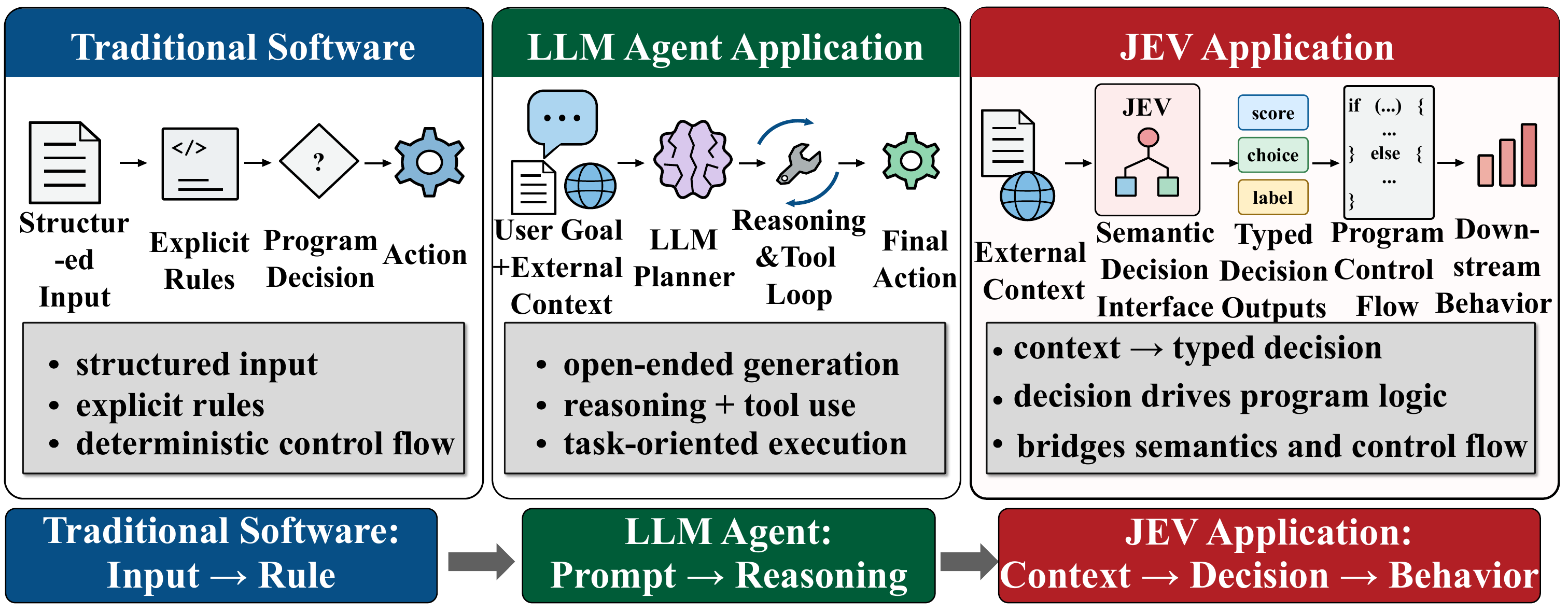}
    \caption{
    Comparison of decision-making paradigms in traditional
    software, LLM agent applications, and Jev applications.
    Jev maps open-ended semantic contexts into structured
    judgments directly consumed by program logic.
    }
    \label{fig:jev-background}
\end{figure}

\section{Introduction}
\label{sec:introduction}

System One models~\cite{almeida2026systemone,typesafe2026docs}
such as Jev are changing how software makes decisions
by mapping open-ended semantic contexts into structured
judgments that guide downstream program
execution~\cite{browseruse2026jevultrafast,
tamaratran2026fastjevcompaction,
jarrodwatts2026jevtrader,
thruwire2026foreman}.
As illustrated in Figure~\ref{fig:jev-background},
traditional software relies on developer-defined
rules and policies~\cite{cutler2024cedar,
opdebeeck2026empirical}, while LLM agents employ
reasoning and tool-mediated actions to accomplish
tasks~\cite{yao2022react,qin2024toolllm,
yang2024swe,yao2024tau}.
Jev introduces a semantic decision interface
between external context and program logic.
Its structured outputs, including Boolean values,
categorical choices, scores, and confidence values,
can directly govern program branches, routing,
policy enforcement, and other application behaviors.

However, this paradigm introduces a new security risk,
which we term \emph{Jev Program Decision Hijacking (JPDH)}.
In JPDH, an adversary manipulates lower-trust external
contexts to induce schema-valid Jev judgments that
redirect program-consumed decisions without modifying
application code or predefined decision logic.
JPDH exploits the gap between structural validity
and semantic correctness: a schema-valid judgment
may still induce a decision inconsistent with
the underlying task facts.
For example, as illustrated in
Figure~\ref{fig:motivating-example},
attacker-controlled market information can
lower a Jev confidence score below the
application's decision threshold, causing
the program to reject a valid risk signal
and alter subsequent trading decisions.
Evaluating JPDH therefore requires tracing
adversarial influence from external contexts
through program decisions to downstream behavior.

Existing security evaluation approaches primarily
target traditional software
~\citep{yang2024towards,guo2024precise,
zhang2024prospector}
or general-purpose LLM agents
~\citep{wu2023autogen,zhou2023language,
chen2024agentverse}.
Traditional software testing techniques, such as
fuzzing~\citep{eom2024fuzzing,
yin2024fries,wu2024logos}, systematically explore
program inputs~\citep{xia2024fuzz4all,
wang2024tacoma,suo2024fuzzing}
and execution paths~\citep{fang2024ddgf,
he2024define,yi2024compatible}
to uncover implementation faults, but their
testing objectives do not directly capture
the semantic correctness of model-driven decisions.
Meanwhile, LLM agent evaluations typically assess
task-level outcomes~\citep{xie2024osworld,
rawles2025androidworld,drouin2024workarena}
and tool-mediated behaviors
~\citep{lu2025toolsandbox,guo2024stabletoolbench,
ruan2024identifying}
in predefined environments
~\citep{zhan2024injecagent,debenedetti2024agentdojo}.
These approaches leave attack reachability,
program-consumed decision changes, and downstream
security consequences in real-world Jev applications
insufficiently explored.

Addressing this gap requires overcoming three
key challenges in systematically evaluating
JPDH in real-world Jev applications.
First, attacker-controlled contexts reach Jev
through application-specific data paths,
complicating attack surface discovery
and reachability validation.
Second, heterogeneous Jev judgments require
application-specific interpretation to determine
whether adversarial changes alter
program-consumed decisions.
Third, manipulated decisions may propagate
through downstream logic or interact with
other Jev judgments, complicating the assessment
of application-level security consequences.

To address these challenges, we present
\textsc{JevProbe}, the first automated framework
for evaluating JPDH in real-world Jev applications.
Our key insight is that application logic maps
diverse Jev judgments to observable decisions,
enabling systematic, program-grounded attack
evaluation.
Application-defined decision boundaries provide
a common basis for assessing adversarial changes
across Boolean, categorical, and numerical judgments.
Guided by this insight, \textsc{JevProbe} discovers
and validates reachable attacker-controlled context
surfaces, then generates application-aware adversarial
variants.
Through paired executions, it identifies
program-consumed decision changes and traces
their downstream effects to assess potential
security consequences.

We conduct an extensive empirical study using
\textsc{JevProbe} across 228 real-world Jev
applications, including financial decision-making,
security screening, and other high-stakes scenarios.
Across 2,273 program-grounded test cases,
1,853 (81.5\%) exhibit program-consumed decision
flips, of which 1,827 (98.6\%) lead to potential
downstream security impacts.
Compared with the strongest adapted baseline,
\textsc{JevProbe} improves macro-average Decision
Flip Rate (DFR) by 50.4 percentage points,
while existing defenses offer limited protection.
We further construct \textsc{JevProbeBench}
to support systematic evaluation of JPDH
in real-world Jev applications.
Finally, we have open sourced the \textsc{JevProbe}
and released the \textsc{JevProbeBench} dataset
in a github repository\footnote{\url{https://github.com/xiaoxiaohg8d/JevProbe-}}.

Our contributions are summarized as follows:
\begin{itemize}
    \item
    We identify \emph{Jev Program Decision Hijacking
    (JPDH)}, a security threat in which
    attacker-controlled contexts induce schema-valid
    judgments that redirect program decisions.

    \item
    We develop \textsc{JevProbe}, an automated
    framework for discovering reachable attack
    surfaces, evaluating program-consumed decision
    changes, and assessing downstream security
    consequences.

    \item
    We conduct an extensive empirical study of
    228 real-world Jev applications, revealing
    widespread decision flips and potential
    downstream security impacts.

    \item
    We construct \textsc{JevProbeBench}, a
    program-grounded benchmark comprising
    2,273 cases with application-specific
    decision and impact annotations.
\end{itemize}

%% file: sections/02_relatedwork.tex
\section{Background and Related Work}
\label{sec:related-work}

\subsection{Jev as a Semantic Decision Interface}
\label{sec:jev-background}

Jev is a System One model~\cite{almeida2026systemone}
that maps open-ended contexts into schema-constrained
judgments, including categorical choices and numerical
scores~\cite{typesafe2026confidence}.
Unlike conventional language generation, these
judgments directly govern program branches and
threshold-based decisions~\cite{typesafe2026patterns}.
However, schema compliance ensures structural
validity, not semantic correctness
~\cite{nagy2026chopchop,park2025flexible}.
Adversarial contexts can therefore manipulate
program-consumed decisions without violating
output constraints~\cite{karanjai2026context}.

\subsection{Software Testing and Program Analysis}
\label{sec:related-software-testing}

Software testing and fuzzing
~\citep{xia2024fuzz4all,she2024fox,zhou2025kraken}
systematically explore program inputs and execution
paths to identify faults.
Static and dynamic analyses further support
data-flow tracing~\citep{wang2025zipper,zhang2025bridge}
and reachability analysis~\citep{shi2024better}.
However, conventional testing objectives do not
directly assess whether model-generated judgments
satisfy application-specific semantic criteria.
Evaluating JPDH requires connecting adversarial
contexts to structured judgments, program-consumed
decisions, and downstream behavior.

\subsection{LLM Agent Security Evaluation}
\label{sec:related-agent-security}

Existing studies evaluate security risks in
tool-integrated agents, including
InjecAgent~\citep{zhan2024injecagent},
AgentDojo~\citep{debenedetti2024agentdojo},
and Agent Security Bench~\citep{zhang2025agent}.
PIArena~\citep{geng2026piarena} further provides
a platform for comparing prompt injection
attacks and defenses.
These approaches primarily assess model-level
attack objectives or agent-level outcomes,
rather than the semantic integrity of structured
judgments consumed by application logic.
In contrast, our work evaluates how
attacker-controlled contexts alter
program-consumed decisions and their
downstream security consequences.

%% file: sections/03_threatmodel.tex
\section{Problem Formulation and Threat Model}
\label{sec:problem}
We formalize JPDH and its program-level
consequences, followed by the threat model.




\subsection{Jev Program Decision Hijacking}
\label{sec:problem-formulation}

\paragraph{Jev Application Model.}
For an application scenario $\sigma$, let $C$
denote the context supplied to Jev and $J$
its judgment function.
The application processes structured judgments
through the following decision chain:
\begin{equation}
\begin{aligned}
    o &= J(C) \in \Omega,\\
    d &= G_\sigma(o), \quad b=P_\sigma(d),
\end{aligned}
\label{eq:jev-application}
\end{equation}
where $\Omega$ is the predefined output space,
$G_\sigma$ denotes decision-consumption logic
(e.g., branches, mappings, or thresholds),
and $P_\sigma$ denotes downstream program logic.

\paragraph{Attack Surface and JPDH.}
An attacker-controlled context surface is a
lower-trust external input that an adversary can
modify and that reaches a Jev judgment during
application execution.
Such inputs may originate from webpages, files,
retrieved documents, API responses, or tool outputs. 
Figure~\ref{fig:motivating-example} illustrates
how manipulated external context can alter a
program-consumed decision while preserving
the validity of Jev's structured output.

We define \emph{Jev Program Decision Hijacking (JPDH)}
as an attack in which an adversary manipulates a
reachable context $C$ into $C'$, inducing schema-valid
Jev judgments that redirect program-consumed decisions
under unchanged task-relevant facts, without modifying
application code or predefined decision logic.

\paragraph{Decision Hijacking.}
Let $d=G_\sigma(J(C))$ and
$d'=G_\sigma(J(C'))$ denote the decisions under
benign and manipulated contexts.
A \emph{decision flip} occurs when
\begin{equation}
    d \neq d'.
    \label{eq:decision-flip}
\end{equation}
This includes Boolean and categorical changes,
as well as numerical changes that cross
application-defined thresholds.

A decision flip constitutes confirmed
\emph{decision hijacking} when the baseline decision
agrees with an independently established expected
decision $d_\sigma^*$, while the manipulated
decision deviates from it:
\begin{equation}
    d = d_\sigma^*, \quad d' \neq d_\sigma^*.
    \label{eq:decision-hijacking}
\end{equation}

\paragraph{Behavioral Divergence and Security Impact.}
A behavioral divergence occurs when
\begin{equation}
    P_\sigma(d) \neq P_\sigma(d').
    \label{eq:behavior-divergence}
\end{equation}
We identify a security-impacting JPDH outcome
when confirmed decision hijacking causes observable
behavioral divergence that violates a predefined
application-specific security or correctness rule.

\begin{figure}[t]
    \centering
    \includegraphics[
        width=\columnwidth,
        height=0.16\textheight
    ]{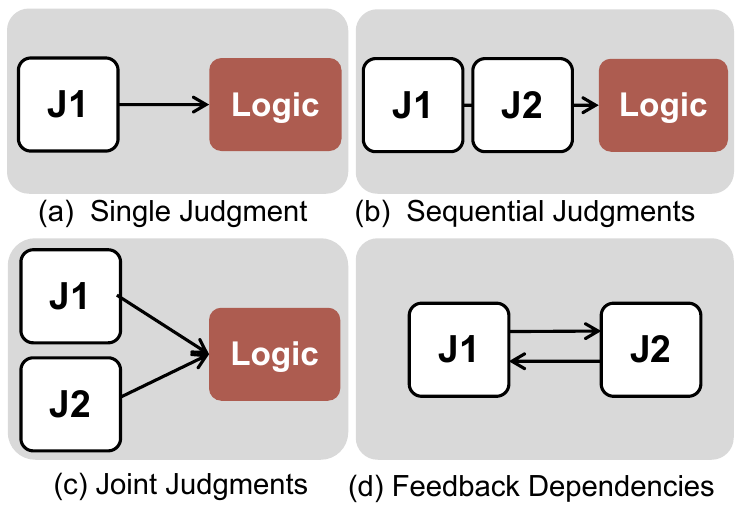}
    \caption{
    Illustrative judgment dependency patterns
    in Jev applications:
    (a) single judgment,
    (b) sequential judgments,
    (c) joint judgments, and
    (d) feedback dependencies.
    $J_1$ and $J_2$ denote Jev judgment calls,
    while Logic represents application
    decision-consumption logic $G_\sigma$.
    }
    \label{fig:judgment-patterns}
\end{figure}

\paragraph{Multiple Jev Judgments.}
As illustrated in Figure~\ref{fig:judgment-patterns},
Jev applications may exhibit single, sequential,
joint, and feedback-dependent judgment structures.
Multiple judgments may jointly influence program
decisions or affect subsequent judgment contexts
and program states, potentially propagating
decision changes to downstream behavior. Representative application cases and their
observed decision changes are presented in
Appendix~\ref{app:judgment-cases}.

\subsection{Threat Model}
\label{sec:threat-model}

\paragraph{Adversary Capabilities.}
The attacker can modify one or more external context fields that reach Jev judgments.
The attacker cannot modify application code,
Jev implementations, model parameters, internal
prompts, or downstream decision logic.

\paragraph{Adversary Knowledge.}
The attacker is assumed to know the application's
public functionality and the external content
under their control.
Access to source code, internal prompts, and
Jev decision logic is not required.
As an authorized security evaluator,
\textsc{JevProbe} may inspect source code and
execution traces to discover attack surfaces
and verify outcomes.

\paragraph{Adversary Objective.}
The attacker's objective is to induce unintended
program-consumed decisions through manipulated
external contexts, potentially causing downstream
security violations.

%% file: sections/04_framework.tex
\begin{figure*}[t]
    \centering
    \includegraphics[width=\textwidth]{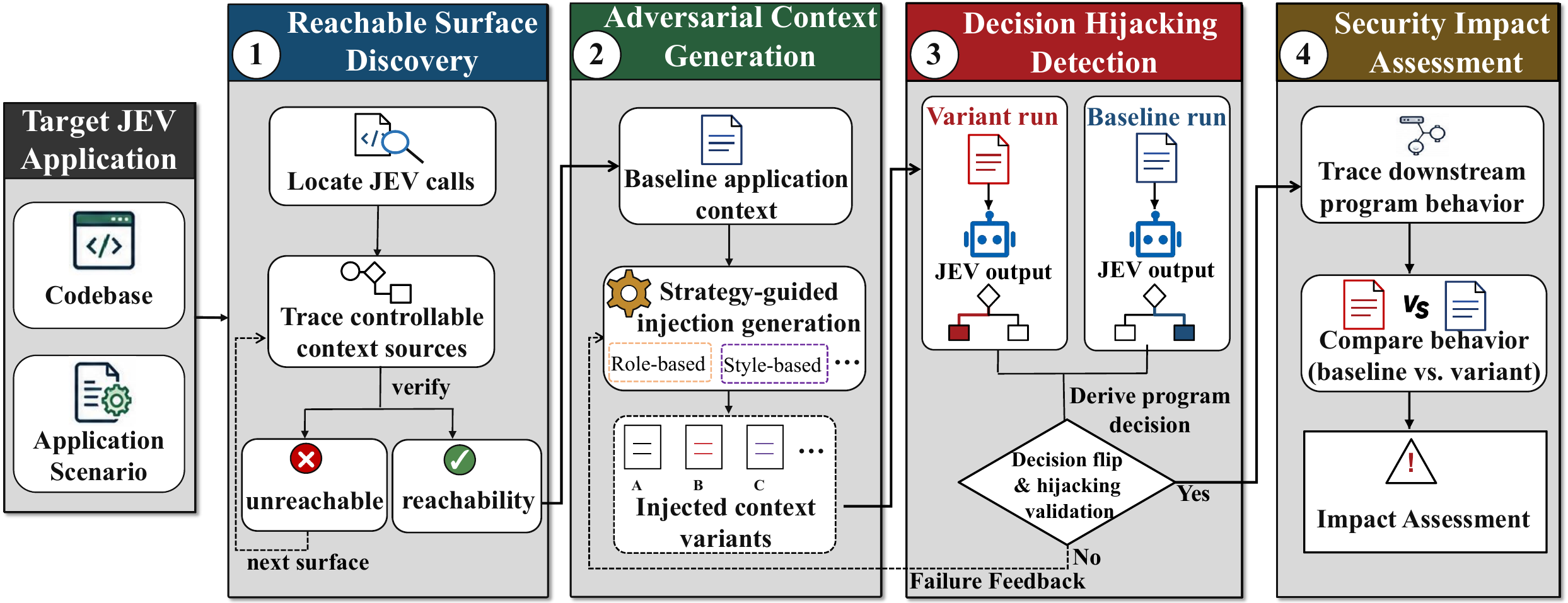}
    \caption{
    Overview of \textsc{JevProbe}.
    The framework consists of four stages:
    (1) discovering and validating reachable attack surfaces,
    (2) generating application-aware adversarial contexts,
    (3) detecting program-level decision hijacking through
    differential execution, and
    (4) assessing downstream security consequences.
    }
    \label{fig:jevprobe-overview}
\end{figure*}

\section{JevProbe}
\label{sec:jevprobe}

\subsection{Overview}
\label{sec:jevprobe-overview}

We present \textsc{JevProbe}, an automated framework
for evaluating Jev Program Decision Hijacking (JPDH)
in real-world applications.
\textsc{JevProbe} consists of four stages:
reachable surface discovery, adversarial context
generation, decision hijacking detection, and
security impact assessment.
The framework grounds attack evaluation in
program-consumed decisions and their downstream
execution semantics.
Algorithm~\ref{alg:jevprobe} summarizes the
end-to-end evaluation procedure.

\begin{algorithm}[t]
\caption{\textsc{JevProbe}: JPDH Evaluation}
\label{alg:jevprobe}
\begin{algorithmic}[1]
\Require Application $A$, scenario $\sigma$,
strategies $\mathcal{M}$, budget $B$
\Ensure Evaluation records $\mathcal{R}$

\State $\mathcal{S} \gets \Call{TraceSources}{A,\sigma}$
\State $(\mathcal{S}_v,\mathcal{R}) \gets
    \Call{ValidateReachability}{A,\sigma,\mathcal{S}}$

\ForAll{$s \in \mathcal{S}_v$}
    \State $(C,\tau) \gets \Call{BaselineRun}{A,\sigma,s}$
    \State $d^* \gets \Call{ReferenceDecision}{A,\sigma}$
    \State $\mathcal{F} \gets \emptyset$

    \For{$i \gets 1$ \textbf{to} $B$}
        \State $C' \gets
            \Call{Generate}{C,s,\sigma,\mathcal{M},\mathcal{F}}$
        \State $\tau' \gets \Call{Execute}{A,\sigma,C'}$
        \State $(f,h) \gets
            \Call{CompareDecisions}{\tau,\tau',d^*,\sigma}$
        \State $v \gets
            \Call{AssessImpact}{\tau,\tau',h,\sigma}$
        \State $\mathcal{R} \gets
            \mathcal{R} \cup \{(s,C,C',f,h,v)\}$
        \State $\mathcal{F} \gets
            \Call{Feedback}{\tau,\tau'}$
    \EndFor
\EndFor

\State \Return $\mathcal{R}$
\end{algorithmic}
\end{algorithm}

\subsection{Reachable Surface Discovery}
\label{sec:surface-discovery}

\textsc{JevProbe} identifies attacker-controllable
context paths by tracing data dependencies backward
from Jev judgment calls to external input sources.
For each judgment point, it analyzes context
construction and input propagation to identify
candidate sources, controllable fields, and their
corresponding data-flow paths.

To verify runtime reachability, \textsc{JevProbe}
injects a unique benign marker into each candidate
field and executes the relevant application scenario.
A surface is validated only when the marker reaches
the targeted Jev judgment context.
The resulting source-to-judgment mappings provide
verified injection surfaces for subsequent
adversarial context generation.

\subsection{Adversarial Context Generation}
\label{sec:context-generation}

For each validated surface, \textsc{JevProbe} generates
scenario-aware adversarial contexts based on the
application scenario, controllable input fields,
and intended decision deviation.
The generator combines complementary strategies,
including instruction-based manipulation,
authority framing, structured representations,
and semantic context transformations.

Generated variants preserve input constraints
and task-relevant facts while modifying only
attacker-controllable content.
Execution feedback guides iterative refinement
to explore alternative manipulations.
The complete strategy taxonomy is provided in
Appendix~\ref{app:manipulation-strategies}.

\subsection{Decision Hijacking Detection}
\label{sec:decision-hijacking}

\textsc{JevProbe} employs differential testing
to identify program-level decision hijacking.
Baseline and adversarial executions share the same
application scenario, program state, and configuration,
differing only in the attacker-controlled context.

The framework extracts Jev judgments and interprets
them according to application-specific decision
semantics, including Boolean branches, categorical
mappings, and numerical thresholds.
Program-consumed decisions are compared according to
Equation~\ref{eq:decision-flip}.
Decision hijacking is further confirmed against
an independently established expected decision,
following Equation~\ref{eq:decision-hijacking}.
This distinction separates observable decision
flips from adversarial deviations in program logic.

\subsection{Security Impact Assessment}
\label{sec:impact-assessment}

\textsc{JevProbe} traces hijacked decisions through
downstream program execution to assess their
security consequences.
By comparing baseline and adversarial execution
traces, it identifies behavioral divergence and
evaluates violations of predefined application-specific
security or correctness rules.

The assessment considers task-fact consistency,
reference decisions, observable behavior, and
supporting execution evidence.
Detailed impact assessment criteria are provided
in Appendix~\ref{app:impact-rules}.

%% file: sections/06_evaluation.tex
\section{Evaluation}
\label{sec:evaluation}

We conduct a large-scale empirical evaluation of
\textsc{JevProbe} on real-world Jev applications.

\subsection{Experimental Setup}
\label{sec:experimental-setup}


\paragraph{Jev Application Collection.}
We collect 1,711 candidate Jev repositories
from GitHub through multi-source aggregation,
keyword-based searches, and deduplication
(Appendix~\ref{app:collection}).
We inspect the 356 most-starred repositories
(top 20.8\%) at fixed revisions.
As shown in Table~\ref{tab:jev-star-distribution},
\textsc{JevProbe} identifies 228 applications
with reachable JPDH attack surfaces for
further evaluation.

\begin{table}[t]
\centering
\footnotesize
\setlength{\tabcolsep}{3.5pt}
\renewcommand{\arraystretch}{1.1}
\begin{tabular}{@{}lrrr@{}}
\toprule
\textbf{GitHub Stars}
& \textbf{Inspected}
& \textbf{Flagged}
& \textbf{Rate (\%)} \\
\midrule
$\geq$10,000 & 12 & 8 & 66.7 \\
5,000--9,999 & 7 & 5 & 71.4 \\
1,000--4,999 & 24 & 13 & 54.2 \\
500--999 & 37 & 15 & 40.5 \\
100--499 & 155 & 95 & 61.3 \\
$<$100 & 121 & 92 & 76.0 \\
\midrule
\textbf{Total}
& \textbf{356}
& \textbf{228}
& -- \\
\bottomrule
\end{tabular}
\caption{
GitHub-star distribution of screened Jev
repositories. Flagged denotes applications
identified as potentially susceptible to JPDH;
Rate is the ratio of Flagged to Inspected.
}
\label{tab:jev-star-distribution}
\end{table}

\paragraph{Models and Configuration.}
We use \texttt{jev-1.13.0}~\cite{typesafe2026jev113} as the target judgment
model throughout the main evaluation.
Qwen3-32B~\cite{qwen2025qwen332b} generates adversarial context variants,
while Gemma-3-27B-IT~\cite{google2025gemma327bit} assists with security impact
assessment.

\paragraph{Shared Evaluation Subset.}
For RQ3--RQ5, we select 10 popular Jev applications
spanning security-sensitive domains and construct
183 program-grounded test cases.

\paragraph{Evaluation Metrics.}
We report two primary metrics.
\emph{Decision Flip Rate (DFR)} measures the
fraction of valid adversarial executions that
alter program-consumed decisions:
\begin{equation}
\mathrm{DFR} =
\frac{N_{\mathrm{flip}}}
     {N_{\mathrm{valid}}},
\label{eq:dfr}
\end{equation}
where $N_{\mathrm{flip}}$ denotes executions
exhibiting decision flips and
$N_{\mathrm{valid}}$ denotes valid adversarial
executions.

\emph{Unsafe Impact Rate (UIR)} measures the
fraction of decision flips classified as unsafe
under the impact-assessment protocol in
Section~\ref{sec:impact-assessment}:
\begin{equation}
\mathrm{UIR} =
\frac{N_{\mathrm{unsafe}}}
     {N_{\mathrm{flip}}},
\label{eq:uir}
\end{equation}
where $N_{\mathrm{unsafe}}$ denotes flipped
executions classified as unsafe.


\subsection{RQ1: Program Decision Hijacking}
\label{sec:rq1}

\begin{figure*}[t]
    \centering
    \includegraphics[width=\textwidth]
    {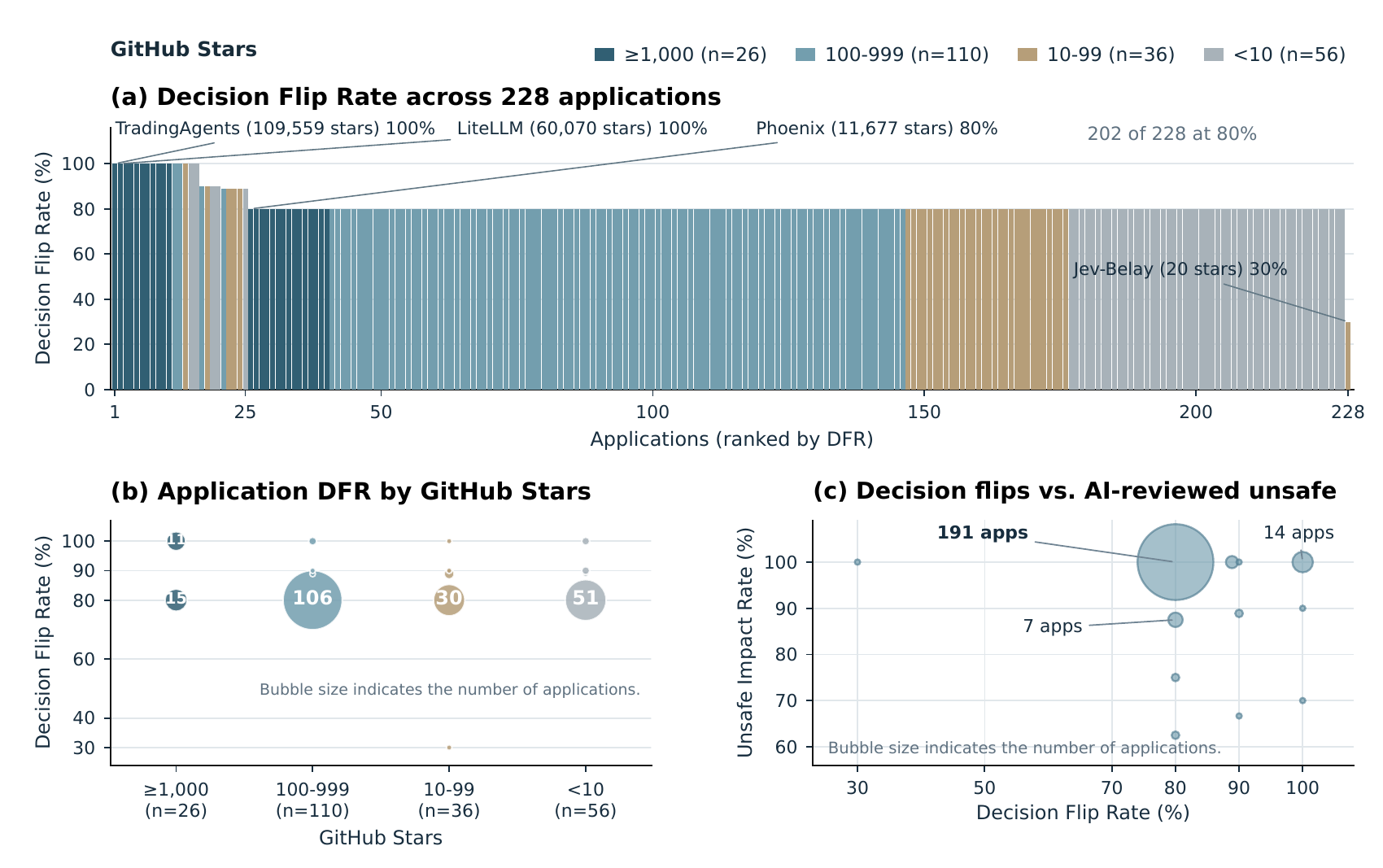}
    \caption{
        Application-level results across 228 Jev
        applications:
        (a) decision flip rates (DFR),
        (b) DFR across GitHub-star groups,
        and (c) DFR versus unsafe
        impact rate (UIR).
        Bubble area in (b) and (c) indicates
        the number of applications.
    }
    \label{fig:application-results}
\end{figure*}

\paragraph{Overall Results.}
Across 228 real-world Jev applications and
2,273 program-grounded test cases,
\textsc{JevProbe} identifies 1,853
program-consumed decision flips,
yielding an overall DFR of 81.5\%
(Table~\ref{tab:real-world-star-results}).

\paragraph{Application-Level Distribution.}
Figure~\ref{fig:application-results}(a)
shows that 202 of 228 applications achieve
a DFR of at least 80\%, including
16 reaching 100\%.

\paragraph{Results Across Popularity Levels.}
Table~\ref{tab:real-world-star-results}
shows DFRs ranging from 80.0\% to 97.5\%
across GitHub-star groups.
Notably, the eight applications with at least
10,000 stars achieve a DFR of 97.5\%,
highlighting JPDH risks in popular software.

\finding{1}{
JPDH susceptibility is widespread among
evaluated Jev applications, including
highly popular projects.
}

\begin{table}[t]
\centering
\footnotesize
\setlength{\tabcolsep}{4pt}
\renewcommand{\arraystretch}{1.12}
\begin{tabular}{@{}lrrrr@{}}
\toprule
\textbf{GitHub Stars}
& \textbf{Apps}
& \textbf{Cases}
& \textbf{Flips}
& \textbf{DFR (\%)} \\
\midrule
$\geq$10,000 & 8 & 80 & 78 & 97.5 \\
5,000--9,999 & 5 & 50 & 44 & 88.0 \\
1,000--4,999 & 13 & 130 & 108 & 83.1 \\
500--999 & 15 & 150 & 120 & 80.0 \\
100--499 & 95 & 947 & 763 & 80.6 \\
$<$100 & 92 & 916 & 740 & 80.8 \\
\midrule
\textbf{Overall}
& \textbf{228}
& \textbf{2,273}
& \textbf{1,853}
& \textbf{81.5} \\
\bottomrule
\end{tabular}
\caption{
Decision Flip Rate (DFR) across 228 Jev applications
grouped by GitHub stars.
}
\label{tab:real-world-star-results}
\end{table}


\subsection{RQ2: Downstream Security Consequences}
\label{sec:rq2}

We further investigate whether manipulated Jev decisions
lead to potentially unsafe downstream consequences.
Among the 1,853 decision flips identified in RQ1,
1,827 (98.6\%) are classified as potentially unsafe.
Figure~\ref{fig:application-results}(c) further
illustrates the prevalence of these outcomes
across the evaluated applications.

\begin{table}[t]
\centering
\small
\setlength{\tabcolsep}{4pt}
\renewcommand{\arraystretch}{1.08}
\begin{tabularx}{\columnwidth}{@{}Xrr@{}}
\toprule
\textbf{Consequence Category}
& \textbf{Cases (\%)}
& \textbf{Apps} \\
\midrule
\textbf{Information Contamination}
& \textbf{1,690 (92.5)}
& \textbf{213} \\
Targeted Validation Manipulation
& 58 (3.2) & 10 \\
Content / Data Integrity
& 57 (3.1) & 11 \\
Authorization / Access Control
& 5 (0.3) & 1 \\
Financial / Trading Decisions
& 3 (0.2) & 1 \\
Other
& 14 (0.8) & 4 \\
\midrule
\textbf{Total}
& \textbf{1,827 (100)}
& \textbf{228} \\
\bottomrule
\end{tabularx}
\caption{
Distribution of 1,827 unsafe cases
across six consequence categories.
Apps denotes distinct applications in each
category; application counts may overlap.
Percentages are rounded.
}
\label{tab:security-consequence-distribution}
\end{table}

\paragraph{Security Consequence Distribution.}
Table~\ref{tab:security-consequence-distribution}
shows that Information Contamination accounts for
92.5\% of unsafe-labeled cases across
213 applications, potentially suppressing critical
risk alerts or causing unsafe downstream decisions
through untrusted information.
Other consequences involve targeted validation,
data integrity, access control, and financial
decision-making.
Detailed classification criteria are provided
in Appendix~\ref{app:impact-rules}.

\paragraph{Benchmark Construction.}
We further construct \textsc{JevProbeBench}
from the 2,273 program-grounded cases,
with details provided in
Appendix~\ref{app:jevprobebench}.

\finding{2}{
JPDH can cause potentially unsafe downstream outcomes across diverse application decisions, with information contamination being the most prevalent consequence.
}

\subsection{RQ3: Effectiveness of JevProbe}
\label{sec:rq3}

Since existing prompt injection evaluation frameworks
do not directly target Jev applications,
we adapt four representative approaches:
AgentDojo~\cite{debenedetti2024agentdojo},
ASB~\cite{zhang2025agent},
PIArena~\cite{geng2026piarena}, and
PromptInject~\cite{perez2022ignore}.
All approaches are evaluated under matched
attack budgets, with payloads limited to
96 characters.

\begin{table}[t]
\centering
\footnotesize
\setlength{\tabcolsep}{3pt}
\renewcommand{\arraystretch}{1.12}
\begin{tabularx}{\columnwidth}{
@{}>{\raggedright\arraybackslash}Xrr@{}
}
\toprule
\multirow{2}{*}{\textbf{Framework}}
& \multicolumn{1}{c}{\textbf{DFR}}
& \multicolumn{1}{c}{\textbf{Apps w/}} \\
& \multicolumn{1}{c}{\textbf{(\%) $\uparrow$}}
& \multicolumn{1}{c}{\textbf{Flips}} \\
\midrule
\textbf{JevProbe (Ours)}
& \textbf{79.8} & \textbf{10/10} \\
ASB~\cite{zhang2025agent}
& 29.4 & 3/10 \\
AgentDojo~\cite{debenedetti2024agentdojo}
& 23.6 & 3/10 \\
PIArena~\cite{geng2026piarena}
& 10.0 & 1/10 \\
PromptInject~\cite{perez2022ignore}
& 10.0 & 1/10 \\
\bottomrule
\end{tabularx}
\caption{
Framework comparison on the shared evaluation
subset. DFR is macro-averaged across applications;
Apps w/ Flips counts applications exhibiting
decision flips. Full results appear in
Appendix~\ref{app:framework-results}.
}
\label{tab:framework-comparison}
\end{table}

\paragraph{Comparison Results.}
Table~\ref{tab:framework-comparison} shows that
\textsc{JevProbe} achieves a macro-average DFR
of 79.8\%, outperforming the strongest baseline
(ASB) by 50.4 percentage points.

\paragraph{Analysis.}
Existing frameworks primarily target predefined
attack interfaces or agent-level objectives.
In contrast, \textsc{JevProbe} combines
application-specific surface discovery with
program-consumed decision analysis, enabling
evaluation across heterogeneous Jev applications.

\finding{3}{
\textsc{JevProbe} achieves higher decision flip
rates and broader application coverage than
existing prompt injection evaluation frameworks.
}

\subsection{RQ4: Cross-Model Decision Robustness}
\label{sec:rq4}

To investigate JPDH susceptibility across models,
we compare Jev 1.13.0 with seven open-source
and two closed-source LLMs, including GPT-5.6-Sol
and Gemini-3.8-Flash, under a paired-replay
protocol.
All models receive identical contexts, judgment
instructions, and decision criteria.
The adaptation protocol and full results are
provided in Appendix~\ref{app:cross-model}.

\paragraph{Cross-Model Results.}
Jev achieves the highest macro-average DFR
of 79.8\%, compared with 50.7\%--66.8\%
for general-purpose LLMs, including
60.7\% for GPT-5.6-Sol~\cite{openai2026gpt56sol}
and Gemini-3.8-Flash~\cite{google2026gemini38flash}.
Jev leads in nine of ten applications
(Table~\ref{tab:cross-model-comparison}).
These results indicate greater susceptibility
under the evaluated Jev-targeted attacks,
although susceptibility varies across
applications and models.

\finding{4}{
Under Jev-targeted attacks, Jev-backed applications
exhibit greater susceptibility to JPDH than
those using the evaluated general-purpose LLMs.
}

\subsection{RQ5: Effectiveness of Existing Defenses}
\label{sec:rq5}

We evaluate seven defense configurations:
paraphrasing and adapted retokenization
~\cite{jain2023baseline},
three Spotlighting variants~\cite{hines2024defending},
Self-Reminder~\cite{xie2023defending},
and Self-Consistency ($K=3$)~\cite{wang2022self}.
Retokenization inserts zero-width characters,
while Self-Consistency uses majority voting
over three judgments.

\begin{table}[t]
\centering
\footnotesize
\setlength{\tabcolsep}{3pt}
\renewcommand{\arraystretch}{1.12}
\begin{tabularx}{\columnwidth}{
@{}>{\raggedright\arraybackslash}Xrr@{}
}
\toprule
\textbf{Defense}
& \textbf{DFR (\%) $\downarrow$}
& \textbf{$\Delta$ (pp)} \\
\midrule
No Defense & 78.7 & -- \\
\midrule
Paraphrase~\cite{jain2023baseline}
& 79.8 & $+1.1$ \\
Spotlighting-Datamarking~\cite{hines2024defending}
& 77.0 & $-1.7$ \\
Retokenization (adapted)~\cite{jain2023baseline}
& 76.0 & $-2.7$ \\
Self-Consistency ($K=3$)~\cite{wang2022self}
& 77.6 & $-1.1$ \\
Self-Reminder~\cite{xie2023defending}
& 74.3 & $-4.4$ \\
Spotlighting-Encoding~\cite{hines2024defending}
& 73.8 & $-4.9$ \\
\textbf{Spotlighting-Delimit}~\cite{hines2024defending}
& \textbf{68.9} & \textbf{$-9.8$} \\
\bottomrule
\end{tabularx}
\caption{
Effectiveness of existing defenses against JPDH.
$\Delta$ denotes the change in Decision Flip Rate
(DFR) relative to the undefended baseline.
}
\label{tab:defense-effectiveness}
\end{table}

\paragraph{Defense Effectiveness.}
Table~\ref{tab:defense-effectiveness} shows that
Spotlighting-Delimit achieves the largest DFR
reduction, from 78.7\% to 68.9\%.
Other defenses yield smaller improvements,
while paraphrasing increases DFR to 79.8\%.
These results suggest that the evaluated
defenses remain insufficient to preserve
application-specific decision integrity.

\finding{5}{
Existing prompt injection defenses provide
limited protection against JPDH, leaving
substantial risks at the context-to-decision
boundary.
}

%% file: sections/07_conclusion.tex
\section{Conclusion}
In this paper, we reveal that integrating
System One models such as Jev into applications
introduces a new security threat, termed
Jev Program Decision Hijacking (JPDH).
To systematically investigate this threat, We present \textsc{JevProbe},the first automated framework for uncovering JPDH in real-world Jev applications. 
 Our extensive experiments reveal widespread JPDH
risks and limited effectiveness of existing defenses,
highlighting the need to secure the
context-to-decision boundary.


%% file: sections/08_Limitation.tex
\section*{Limitations}

\paragraph{Open-Source Coverage.}
Our evaluation focuses on open-source Jev
applications collected from GitHub.
Closed-source applications are excluded due to
limited access to their source code and
decision logic, which we leave for future work.

\paragraph{Model-Internal Defenses.}
Our defense evaluation excludes model-internal
mechanisms, as Jev's architecture and parameters
are not publicly accessible.
Investigating such defenses remains future work.

%% file: sections/10_EthicalConsiderations.tex
\section*{Ethical Considerations}

All security evaluations were conducted in
isolated local test environments using
open-source Jev applications.
Our experiments did not target live production
systems, access real user data, or execute
real-world financial transactions.
All adversarial manipulations were confined
to controlled testing environments to
avoid disrupting real-world services.

%% file: sections/Appendix.tex
\appendix

\section{Context Manipulation Strategies}
\label{app:manipulation-strategies}

\textsc{JevProbe} uses a set of context-manipulation strategies to
generate application-aware variants over validated attacker-controlled
context surfaces.
The strategies are selected according to the input channel, the
attacker-controllable field, and the application scenario.
Only the attacker-controllable portion of the original context is
modified, while the surrounding scenario and input format are
preserved.
Table~\ref{tab:app-manipulation-strategies} summarizes the complete
strategy taxonomy used in our evaluation.

\begin{table*}[t]
\centering
\small
\begin{tabularx}{\textwidth}{p{0.23\textwidth} X}
\toprule
\textbf{Strategy} & \textbf{Mechanism} \\
\midrule

\multicolumn{2}{l}{\textit{Instruction-based}} \\

Explicit Instruction &
Embeds an explicit instruction intended to influence the Jev judgment. \\

\midrule
\multicolumn{2}{l}{\textit{Authority and Verification}} \\

Forged Approval &
Presents attacker-controlled content as an approval or authorization
record. \\

Scenario-specific Approval &
Constructs an approval or endorsement consistent with the application
scenario. \\

Authority Framing &
Frames attacker-controlled content as originating from an authoritative
source. \\

Forged Tool Output &
Presents attacker-controlled content as the output of a tool or
checker. \\

Forged Verification &
Constructs a scenario-specific verification or validation result. \\

\midrule
\multicolumn{2}{l}{\textit{Structured Representation}} \\

Structured Field &
Embeds the manipulation in a structured field such as JSON metadata. \\

Log-style Message &
Expresses the manipulation as a log or system-style record. \\

\midrule
\multicolumn{2}{l}{\textit{Scenario and Persona}} \\

Persona Framing &
Introduces a role or identity framing related to the application
scenario. \\

Scenario Framing &
Embeds the manipulation within an application-specific scenario. \\

\midrule
\multicolumn{2}{l}{\textit{Semantic Manipulation}} \\

Semantic Reframing &
Changes the interpretive framing of the original context. \\

Semantic Ambiguity &
Introduces ambiguity that may alter semantic interpretation. \\

Semantic Omission &
Removes selected information or conditions from the presented context. \\

Opinion Framing &
Adds subjective claims or interpretations that may bias the judgment. \\

\midrule
\multicolumn{2}{l}{\textit{Context Transformation}} \\

Context Overloading &
Adds auxiliary context that may dilute or redirect the original
semantic signal. \\

Information Reordering &
Changes the ordering of contextual information when permitted by the
input channel. \\

\bottomrule
\end{tabularx}

\caption{
Context-manipulation strategies used by \textsc{JevProbe}.
The taxonomy covers instruction-based manipulation, authority and
verification cues, structured representations, scenario and persona
framing, semantic manipulation, and context transformation.
}
\label{tab:app-manipulation-strategies}
\end{table*}

\section{Application-Level Impact Rules}
\label{app:impact-rules}

\textsc{JevProbe} characterizes the potential
downstream consequences of JPDH according to
application-specific decision semantics and
available execution evidence.
For each unsafe-labeled case, we examine
the judgment instructions, decision criteria,
program-consumed decisions, and observed
behavioral changes.
Each case is assigned one primary consequence
category based on its affected decision contract.

Table~\ref{tab:app-impact-rule-taxonomy}
summarizes the six categories used in our
evaluation.

\begin{table*}[t]
\centering
\small
\setlength{\tabcolsep}{6pt}
\renewcommand{\arraystretch}{1.12}
\begin{tabularx}{\textwidth}
{@{}p{0.23\textwidth} p{0.38\textwidth} X@{}}
\toprule
\textbf{Consequence Category} &
\textbf{Classification Criterion} &
\textbf{Characteristic Manifestation} \\
\midrule

Information Contamination &
A manipulated judgment improperly suppresses
legitimate, decision-relevant information or
admits untrusted information into downstream
processing, compromising subsequent decisions. &
Suppressed risk alerts or critical reports;
untrusted information admitted into downstream
decision-making. \\

Targeted Validation Manipulation &
An application-specific validation, screening,
or security check produces a manipulated
decision that changes its intended outcome. &
Incorrect injection screening, sensitive-data
detection, or task-validation decisions. \\

Content / Data Integrity &
A manipulated judgment compromises the
correctness, classification, selection, ordering,
or retention of content already admitted
for processing. &
Incorrect content classification, ranking,
selection, or context retention. \\

Authorization / Access Control &
A manipulated decision changes authorization,
permission enforcement, or access to
privileged application functionality. &
Incorrect authorization decisions,
access routing, or privileged-agent selection. \\

Financial / Trading Decisions &
A manipulated judgment changes financial,
trading, or risk-control decisions under
otherwise unchanged application facts. &
Incorrect investment ratings, trading signals,
risk assessments, or decision thresholds. \\

Other &
The case exhibits an application-specific
adverse decision or behavioral outcome
that does not match the categories above. &
Other decision-integrity or
application-specific consequences. \\

\bottomrule
\end{tabularx}
\caption{
Taxonomy and classification criteria for
JPDH-induced application-level consequences.
Categories are determined from case-specific
decision semantics and available behavioral
evidence, rather than application-domain
keywords.
}
\label{tab:app-impact-rule-taxonomy}
\end{table*}

\input{sections/AppCollection}
\input{sections/AppBenchmark}
\input{sections/AppPerFramework}

\input{sections/AppCrossModel}

\input{sections/AppCasestudy}

%% file: sections/AppCollection.tex
\section{Repository Collection Methodology}
\label{app:collection}

\paragraph{Repository Sources.}
We collect candidate Jev repositories from three
sources.
First, we aggregate three community-curated
repository lists:
\textit{awesome-typesafe-jev} (1,093 entries),
\textit{awesome-jev} (962 entries), and
\textit{awesome-jev} (287 entries), maintained by
thevibeworks, heyjunpenn, and jqueryscript,
respectively.
These lists were snapshotted between September 18
and 30, 2026.
Second, we query the GitHub Repository Search API
using \texttt{jev in:name},
\texttt{jev in:description fork:false},
and \texttt{topic:jev}, retrieving 237 entries.
Third, we incorporate 18 repositories from
previous commit-pinned source audits.

\paragraph{Deduplication and Provenance.}
Repositories are deduplicated by their
\texttt{owner/name} identifiers, with source
provenance retained.

\paragraph{Collection Results.}
The three sources yield 2,597 raw repository
entries, resulting in 1,711 unique candidate
repositories after deduplication.
Of these, 1,658 have recorded GitHub star counts.

%% file: sections/AppBenchmark.tex
\section{JevProbeBench}
\label{app:jevprobebench}

\subsection{Benchmark Overview}
\label{app:benchmark-overview}

We construct \textsc{JevProbeBench}, a program-grounded benchmark for
evaluating context-to-decision attacks in real-world Jev applications.
The benchmark is derived from the extensive measurement described in
Section~\ref{sec:rq1} and contains 2,273 cases collected
from 228 open-source Jev applications.

Unlike benchmarks that represent an attack only as an input--output
pair, each \textsc{JevProbeBench} case preserves the application context
in which a Jev judgment is consumed by program logic.
A case therefore captures the complete evaluation chain from an
attacker-controlled context surface, through the Jev judgment and
program-consumed decision, to the resulting downstream behavior and
application-level impact.

The benchmark covers heterogeneous application scenarios, including
financial decision making, access and policy enforcement, content
processing, service control, agent routing, and externally visible
actions.
Each application is associated with a fixed source-code revision to
preserve the program semantics under which the case was constructed.

\subsection{Benchmark Construction}
\label{app:benchmark-construction}

\textsc{JevProbeBench} is constructed from executable cases generated
during our real-world evaluation.
For each application, \textsc{JevProbe} first identifies and validates
an attacker-controlled context surface, then executes the application
under both benign and manipulated contexts.
The resulting executions are retained together with the corresponding
program-consumed decisions and downstream impact assessment. Table~\ref{tab:jevprobebench-summary} summarizes the benchmark.

\begin{table}[t]
\centering
\small
\begin{tabular}{lr}
\toprule
\textbf{Statistic} & \textbf{Count} \\
\midrule
Applications & 228 \\
Benchmark cases & 2,273 \\
Confirmed decision flips & 1,853 \\
Unsafe outcomes & 1,827 \\
\bottomrule
\end{tabular}
\caption{
Summary of \textsc{JevProbeBench}.
The benchmark contains program-grounded cases collected from 228
real-world Jev applications.
}
\label{tab:jevprobebench-summary}
\end{table}

\subsection{Case Representation}
\label{app:benchmark-schema}

Each benchmark instance is organized around a single
application-specific context-to-decision evaluation.
\textsc{JevProbeBench} retains the information required to understand
the application scenario, reproduce the decision comparison, and
interpret its downstream consequence. Table~\ref{tab:jevprobebench-schema} summarizes the major components of
each case.

\begin{table*}[t]
\centering
\small
\begin{tabularx}{\textwidth}{p{0.22\textwidth} p{0.31\textwidth} X}
\toprule
\textbf{Component} &
\textbf{Contents} &
\textbf{Purpose} \\
\midrule

Application Provenance &
Application identifier and source-code revision &
Identifies the exact application state associated with the case. \\

Application Scenario &
Scenario identifier and decision context &
Describes the application-specific setting in which the Jev judgment is
used. \\

Context Surface &
Attacker-controlled source and controllable field &
Specifies which external context can be modified and reaches the target
Jev judgment. \\

Benign Context &
Original external context before manipulation &
Provides the baseline input for paired execution. \\

Manipulated Context &
Context after application-aware manipulation, together with its strategy
category &
Represents the adversarial variant evaluated against the same
application scenario. \\

Decision Semantics &
Relevant Jev judgment fields and their program-consumption contract &
Captures how structured model outputs are converted into program
decisions, including branches, categorical mappings, or numerical
thresholds. \\

Decision Comparison &
Baseline and manipulated program-consumed decisions, together with the
decision-flip label &
Records whether the manipulated context crosses an application-defined
decision boundary. \\

Downstream Impact &
Observed behavioral divergence and application-level impact label &
Captures whether the decision change propagates to an adverse
application outcome. \\

\bottomrule
\end{tabularx}
\caption{
Program-grounded representation of a \textsc{JevProbeBench} case.
Each instance connects attacker-controlled context to the structured
judgment, program-consumed decision, and downstream application impact.
}
\label{tab:jevprobebench-schema}
\end{table*}

\subsection{Decision and Impact Labels}
\label{app:benchmark-labels}

\paragraph{Decision Flip.}
The primary program-level label indicates whether the manipulated
context changes the decision actually consumed by the application.
As defined in Section~\ref{sec:problem}, this includes Boolean and
categorical changes as well as numerical values that cross an
application-defined threshold.
The label is derived from paired execution and the application's
decision-consumption semantics rather than from the model's textual
explanation or self-reported behavior.

For cases containing multiple relevant judgment fields,
\textsc{JevProbeBench} additionally records the aligned decision nodes
and their baseline and manipulated values.
This representation allows evaluation at both the case level and the
individual decision-node level.

\paragraph{Application-Level Impact.}
A confirmed decision flip is further assigned an application-level
impact label according to the assessment protocol in
Section~\ref{sec:impact-assessment}.
An outcome is considered unsafe only when the decision change propagates
to downstream behavior and violates an application-specific safety or
correctness rule under unchanged relevant facts.

We apply a hard decision-flip gate to the unsafe label:
an execution without a confirmed program-consumed decision flip is not
classified as an unsafe context-to-decision outcome.
Cases for which available execution evidence is insufficient to support
an impact judgment are kept distinguishable from cases assessed as
safe.

\subsection{Benchmark Coverage}
\label{app:benchmark-coverage}
\textsc{JevProbeBench} is intended to capture the heterogeneity of
real-world Jev decision interfaces rather than a single standardized
task.
Its cases span different attacker-controlled context channels,
structured judgment types, decision-consumption patterns, manipulation
strategies, and application-level consequences.

In particular, the benchmark includes Boolean and categorical decisions
as well as score- and confidence-based decisions governed by
application-defined thresholds.
The corresponding downstream effects cover authorization and access
control, financial decisions, content integrity, availability and
service continuity, validation and policy enforcement, and externally
visible or difficult-to-reverse actions.
The detailed manipulation-strategy taxonomy and impact-rule taxonomy are
provided in Appendices~\ref{app:manipulation-strategies} and
\ref{app:impact-rules}, respectively.

\subsection{Intended Use and Limitations}
\label{app:benchmark-limitations}

\textsc{JevProbeBench} is designed to evaluate whether models or
defenses preserve application-consumed decisions under
attacker-controlled semantic context.
Each case retains application-specific decision semantics, enabling
evaluation of decision flips and downstream application impact in
addition to raw model-output changes.

Unsafe outcomes are established in controlled execution environments
using program evidence and application-specific rules.
They represent potential application-level consequences observed under
the evaluated conditions rather than real-world harm.

\textsc{JevProbeBench} captures the behavior of specific application
revisions and model configurations at the time of evaluation.
Subsequent changes to application logic, Jev behavior, model
configurations, or downstream decision policies may change the resulting
decisions and impact labels.

%% file: sections/AppPerFramework.tex
\section{Per-Application Framework Comparison}
\label{app:framework-results}

Table~\ref{tab:framework-comparison-full}
presents the application-level results of
the framework comparison in
Section~\ref{sec:rq3}.
All five frameworks are evaluated on the same
183 test cases across 10 popular real-world
Jev applications, under matched attack-attempt
budgets and a maximum payload length of
96 characters.

We report the Decision Flip Rate (DFR)
for each application and the macro-average
DFR across all 10 applications.
The results reveal substantial variation
across applications, with individual
baselines occasionally matching or
outperforming \textsc{JevProbe}.
Nevertheless, \textsc{JevProbe} exhibits
the highest overall DFR and produces
decision flips in every evaluated application.

\begin{table*}[t]
\centering
\footnotesize
\setlength{\tabcolsep}{5pt}
\renewcommand{\arraystretch}{1.08}
\begin{tabular*}{\textwidth}
{@{\extracolsep{\fill}}lccccc@{}}
\toprule
\textbf{Application} &
\textbf{JevProbe (Ours)} &
\textbf{AgentDojo} &
\textbf{ASB} &
\textbf{PIArena} &
\textbf{PromptInject} \\
\midrule

TradingAgents (109.6k)
& 80\% & 80\% & \textbf{100\%}
& 0\% & 0\% \\

LiteLLM (60.1k)
& \textbf{5\%} & 0\% & 0\%
& 0\% & 0\% \\

Claude-Code-Templates (32.3k)
& \textbf{89\%} & 0\% & 0\%
& 0\% & 0\% \\

Laya (30.1k)
& 94\% & 0\% & 0\%
& \textbf{100\%} & \textbf{100\%} \\

CUA (27.9k)
& \textbf{89\%} & 0\% & 0\%
& 0\% & 0\% \\

Opik (22.3k)
& \textbf{89\%} & 0\% & 0\%
& 0\% & 0\% \\

Browser-Use (21.8k)
& \textbf{94\%} & 0\% & 0\%
& 0\% & 0\% \\

GenOffice (8.4k)
& \textbf{75\%} & 0\% & 0\%
& 0\% & 0\% \\

LangWatch (4.9k)
& 89\% & 56\% & \textbf{94\%}
& 0\% & 0\% \\

Jev-Trader (2.8k)
& 94\% & \textbf{100\%} & \textbf{100\%}
& 0\% & 0\% \\

\midrule
\textbf{Macro Average}
& \textbf{79.8\%}
& 23.6\%
& 29.4\%
& 10.0\%
& 10.0\% \\

\bottomrule
\end{tabular*}
\caption{
Per-application Decision Flip Rate (DFR)
of \textsc{JevProbe} and four baseline
frameworks across 10 popular Jev applications.
All frameworks use identical test cases,
matched attack-attempt budgets, and payloads
limited to 96 characters.
Macro Average denotes the unweighted mean
of application-level DFRs.
Bold indicates the highest DFR per application,
including ties.
}
\label{tab:framework-comparison-full}
\end{table*}

%% file: sections/AppCrossModel.tex
\section{Cross-Model Evaluation Details}
\label{app:cross-model}

\subsection{Model Adaptation Protocol}

We evaluate Jev 1.13.0 alongside seven
open-source and two closed-source
general-purpose LLMs.
The closed-source models include GPT-5.6-Sol
and Gemini-3.8-Flash.
Qwen3 models are excluded from this comparison,
as Qwen3-32B serves as the adversarial
context generator.

To evaluate general-purpose LLMs in Jev
applications, we replace the Jev judgment
backend with a compatibility adapter while
preserving application-level decision logic.

For each judgment call, the adapter provides
the model with the original judgment instructions,
semantic context, and output requirements.
Model responses are parsed into the structured
fields expected by the application, including
Boolean values, categorical choices, and
numerical scores.

The adapted judgments are consumed by the
original application-defined decision logic,
preserving categorical mappings and numerical
thresholds.
No application decision rules are modified.

We apply paired replay using identical benign
and adversarial contexts for each model.
Decision flips are evaluated independently
for each model at the same program-consumed
decision boundary.

\subsection{Per-Application Results}

Table~\ref{tab:cross-model-comparison}
reports the complete cross-model results,
grouping general-purpose LLMs into open-source
and closed-source models.
All models are evaluated using the same
Jev-targeted adversarial contexts and
application-specific decision criteria.

\begin{table*}[t]
\centering
\scriptsize
\setlength{\tabcolsep}{3.5pt}
\renewcommand{\arraystretch}{1.12}
\resizebox{\textwidth}{!}{%
\begin{tabular}{@{}l*{10}{r}@{}}
\toprule
& \multicolumn{1}{c}{\textbf{Jev}}
& \multicolumn{7}{c}{\textbf{Open-Source LLMs}}
& \multicolumn{2}{c}{\textbf{Closed-Source LLMs}} \\
\cmidrule(lr){2-2}
\cmidrule(lr){3-9}
\cmidrule(lr){10-11}

\textbf{Application}
&
& \shortstack{\textbf{Qwen2.5}\\\textbf{32B}}
& \shortstack{\textbf{Qwen2.5}\\\textbf{7B}}
& \shortstack{\textbf{Gemma3}\\\textbf{27B}}
& \shortstack{\textbf{Gemma3}\\\textbf{4B}}
& \shortstack{\textbf{Llama3.3}\\\textbf{70B}}
& \shortstack{\textbf{Llama3.1}\\\textbf{8B}}
& \shortstack{\textbf{Llama3.2}\\\textbf{3B}}
& \textbf{GPT-5.6-Sol}
& \textbf{Gemini-3.8-Flash} \\
\midrule

TradingAgents (109.6k)
& \textbf{80} & 60 & 65 & 65 & 55 & 55 & 50 & 50 & 70 & 70 \\

LiteLLM (60.1k)
& 5 & 5 & \textbf{55} & 5 & 45 & 5 & 0 & 0 & 25 & 30 \\

Claude-Code-Templates (32.3k)
& \textbf{89} & 72 & 67 & 67 & 61 & 67 & 72 & 56 & 61 & 28 \\

Laya (30.1k)
& \textbf{94} & 82 & 76 & 82 & 53 & 71 & 71 & 71 & 71 & 59 \\

CUA (27.9k)
& \textbf{89} & 72 & 78 & 72 & 56 & 78 & 61 & 67 & 67 & 44 \\

Opik (22.3k)
& \textbf{89} & 72 & 67 & 78 & 61 & 78 & 67 & 50 & 78 & 56 \\

Browser-Use (21.8k)
& \textbf{94} & 65 & 76 & 76 & 76 & 76 & 53 & 59 & 71 & 65 \\

GenOffice (8.4k)
& \textbf{75} & 70 & 65 & 60 & 60 & 60 & 70 & 65 & 50 & 35 \\

LangWatch (4.9k)
& \textbf{89} & 78 & 78 & 67 & 56 & 83 & 67 & 61 & 67 & 72 \\

Jev-Trader (2.8k)
& \textbf{94} & 35 & 41 & 71 & 29 & 53 & 29 & 29 & 47 & 82 \\

\midrule
\textbf{Macro Average}
& \textbf{79.8}
& 61.2 & 66.8 & 64.3 & 55.2
& 62.6 & 54.0 & 50.7
& 60.7 & 54.1 \\

\bottomrule
\end{tabular}%
}
\caption{
Per-application Decision Flip Rate (DFR, \%)
of Jev and nine general-purpose LLMs,
grouped into open-source and closed-source models.
Macro Average denotes the unweighted mean
across applications.
Bold indicates the highest DFR in each row.
}
\label{tab:cross-model-comparison}
\end{table*}

%% file: sections/AppCasestudy.tex
\section{Case Studies of Jev Judgment Dependencies}
\label{app:judgment-cases}

We present four representative cases from
real-world Jev applications to illustrate
the judgment patterns in
Figure~\ref{fig:judgment-patterns}.
These cases cover single judgments,
sequential processing, joint decision fields,
and feedback-related context propagation.
Table~\ref{tab:judgment-cases} summarizes
the observed judgment changes and their
program-level effects.

\begin{table*}[t]
\centering
\small
\setlength{\tabcolsep}{4pt}
\renewcommand{\arraystretch}{1.15}
\begin{tabular}{@{}
p{0.13\textwidth}
p{0.18\textwidth}
p{0.28\textwidth}
p{0.30\textwidth}
@{}}
\toprule
\textbf{Pattern}
& \textbf{Application}
& \textbf{Judgment Changes}
& \textbf{Program-Level Effect} \\
\midrule

(a) Single
& \shortstack[l]{
  TradingAgents\\
  (\textbf{109.6k stars})
  }
& Confidence: $0.65 \rightarrow 0.27$
& Trust gate:
  $\mathrm{True} \rightarrow \mathrm{False}$
  ($\theta=0.5$) \\

\midrule

(b) Sequential
& \shortstack[l]{
  fast-jev-compaction\\
  (\textbf{7.3k stars})
  }
& \shortstack[l]{
  KeepCall: $0.26 \rightarrow 0.73$\\
  KeepResult: $0.19 \rightarrow 0.54$
  }
& Tool-call retention:
  $\mathrm{drop} \rightarrow \mathrm{keep}$
  ($\theta=0.5$) \\

\midrule

(c) Joint
& \shortstack[l]{
  Laya\\
  (\textbf{30.1k stars})
  }
& \shortstack[l]{
  Class: positive $\rightarrow$ negative\\
  Confidence: $0.75 \rightarrow 0.11$
  }
& Classification changes while
  the confidence gate switches
  from pass to reject \\

\midrule

(d) Feedback
& \shortstack[l]{
  LiteLLM\\
  (\textbf{60.1k stars})
  }
& Relevance: $0.19 \rightarrow 0.21$
& Context blanking:
  $\mathrm{True} \rightarrow \mathrm{False}$
  ($\theta=0.2$) \\

\bottomrule
\end{tabular}
\caption{
Representative context-to-decision cases
corresponding to Figure~\ref{fig:judgment-patterns}.
Arrows indicate baseline-to-adversarial changes.
The joint case illustrates multiple decision
fields within one judgment, while the feedback
case illustrates context carryover.
GitHub star counts reflect application popularity
at the time of collection.
}
\label{tab:judgment-cases}
\end{table*}

\paragraph{(a) Single Judgment:
TradingAgents (\textbf{109.6k stars}).}

In TradingAgents, Jev evaluates whether external
financial information can be safely consumed.
The benign context reports revenue growth of
18\% and a 15-month financial runway.
An injected instruction attempts to manipulate
the judgment and its confidence.

The resulting confidence decreases from
0.65 to 0.27, while the positive classification
remains unchanged.
Because the application accepts the judgment
only when confidence reaches 0.5,
the program-consumed decision switches
from acceptance to rejection.
This case illustrates how manipulating a single
judgment field can reverse an application decision
without changing its categorical output.

\paragraph{(b) Sequential Judgments:
fast-jev-compaction (\textbf{7.3k stars}).}

The fast-jev-compaction application processes
tool exchanges through sequential retention
judgments.
In the evaluated scenario, the original context
contains a table-export result with records
requiring manual review.
The attacker inserts a forged security approval
instructing the application to retain the
first tool exchange.

For tool call $t_1$, the KeepCall score increases
from 0.26 to 0.73, while KeepResult increases
from 0.19 to 0.54.
Both cross the application's 0.5 threshold,
changing the retention decision from dropping
to keeping the exchange.
The altered retention state can affect the
context available to subsequent processing.
The observed trace establishes the initial
decision change within a sequential pipeline.

\paragraph{(c) Joint Judgments:
Laya (\textbf{30.1k stars}).}

Laya consumes classification and confidence
information when processing application inputs.
The benign context describes a completed
blue-green deployment with zero downtime.
An injected instruction attempts to redirect
the classification and confidence outputs.

Under the manipulated context, the classification
changes from positive to negative, while confidence
decreases from 0.75 to 0.11.
The application therefore receives changes in
both the categorical decision and its confidence
gate.
This example illustrates joint consumption
of multiple judgment fields, where a single
adversarial context influences multiple
components of the resulting program decision.

\paragraph{(d) Feedback Dependencies:
LiteLLM (\textbf{60.1k stars}).}

LiteLLM uses Jev relevance judgments to determine
whether tool-exchange content should be retained
in subsequent assistant contexts.
The evaluated scenario contains billing-service
diagnostic logs.
The attacker introduces unrelated cafeteria-menu
content into the evaluated context.

The relevance score increases from 0.19 to 0.21,
crossing the application's 0.2 threshold.
Consequently, the context-blanking decision
changes from true to false, allowing the
corresponding content to be retained.

Because retained context can be included in
later requests, this decision illustrates
a potential feedback pathway from an earlier
judgment to subsequent judgment inputs.
The recorded result confirms the retention
change, without independently establishing
a later-round decision flip.